# Solving the "magic angle" challenge in determining molecular orientation heterogeneity at interfaces[†]

Zhiguo Li,[a‡] Jiaxi Wang,[a‡] Yingmin Li,[b] Wei Xiong[*a,b]

We introduce a novel method to determine the orientation heterogeneity (mean tilt angle and orientational distribution) of molecules at interfaces using heterodyne two-dimensional sum frequency generation spectroscopy. By doing so, we not only have solved the long-standing "magic angle" challenge, i.e. the measurement of molecular orientation by assuming a narrow orientational distribution results in ambiguities, but we also are able to determine the orientational distribution, which is otherwise difficult to measure. We applied our new method to a $CO_2$ reduction catalyst/gold interface and found that the catalysts formed a monolayer with a mean tilt angle between the quasi-$C_3$ symmetric axis of the catalysts and the surface normal of 53°, with 5° orientational distribution. Although applied to a specific system, this method is a general way to determine the orientation heterogeneity of an ensemble-averaged molecular interface, which can potentially be applied to a wide-range of energy material, catalytic and biological interfaces.

## Introduction

Molecular orientation at interfaces in condense phase systems, including heterogeneous catalysts,[1–3] energy materials[4] and biological membranes,[5,6] is often heterogeneous. It can be an ordered monolayer with all molecules tilted in the same distribution, monolayers with randomly orientated molecules, or monolayers with complicated orientational distributions. Understanding the complex molecular orientation at these interfaces is critical to reveal the surface structure-function relationships. For instance, the orientation of adsorbed acrolein on the surface of Ag(111) affects the proximity C=C bond of reactants to the surface, which has been shown to influence both the hydrogenation reaction's activity and selectivity.[1] Another example is that in biological lipid membranes, lipids adopt new orientations when antimicrobial peptides intrude into the lipid membranes. Measuring the change of lipid orientation can elucidate the microscopic pictures of how the membrane morphologies respond to the invading peptides,[6] which could have important implications for drug designs.

A common model used to describe the molecular orientation of monolayers is a Gaussian distribution. Hence, at least two physical quantities need to be measured – the mean tilt angle $\theta_0$ and the orientational distribution $\sigma$, which we refer to as the ($\theta_0$, $\sigma$) pair or orientation heterogeneity hereafter. We note more complex orientational distributions could exist. In this report, we focus on the Gaussian distribution model, which can provide a basic description of the orientation heterogeneity of the molecular monolayers.

It is a challenge to measure orientation heterogeneity, even with the Gaussian distribution model. For decades, surface-specific vibrational sum frequency generation spectroscopy (referred to as 1D VSFG hereafter)[7–15] has been applied to determine the mean tilt angle, under the assumption of a narrow orientational distribution. However, by assuming a narrow angular distribution, the orientational distribution knowledge is lost, and the measured mean tilt angle can deviate from the real mean tilt angle when the orientational distribution is large, which is the well-known "magic angle" challenge. This issue arises when 1D VSFG measures an orientation parameter $D_1$[16–18] that is equal to

$$D_1 = <\cos^3\theta>/<\cos\theta> = \int_0^\pi \cos^3\theta * G(\theta;\theta_0,\sigma) * d\theta / \int_0^\pi \cos\theta * G(\theta;\theta_0,\sigma) * d\theta \qquad (1)$$

where $\theta$ is the angle between the surface normal and the molecular axis, and the bracket means orientational average (integrating the products of cosine functions and a modified Gaussian distribution $G(\theta;\theta_0,\sigma)$ from 0° to 180°).[16,19–26] Clearly, $D_1$ is a function of the ($\theta_0$, $\sigma$) pair, so a single $D_1$ cannot warrant a unique solution to $\theta_0$ and $\sigma$.

When the orientational distribution is assumed to be narrow ($\sigma$=0°), $D_1$ can be simplified to $\cos^2<\theta_0>$, and the mean tilt angle ($\theta_0$) can be calculated. While it works fine for well-ordered, self-assembled monolayers, for most molecular monolayers, this narrow angular distribution assumption is not always valid.[27] In the broad angular distribution case, each $D_1$ corresponds to many pairs of ($\theta_0$, $\sigma$)[28–32] (solid lines in Fig. 1a). For instance, if $D_1$=0.600 is measured, the net orientation can either be $\theta_0$=39.2° with a uniform distribution ($\sigma$=0°) or any other mean tilt angle with broad distribution ($\sigma$=90°). (See Fig. 1b.) This 39.2°, referred to as the "magic angle", represents the extreme case of ambiguities in determining orientation heterogeneity using 1D VSFG,[27] while similar ambiguity also remains for any other $D_1$ values. In summary, the "magic angle" challenge results in two areas of uncertainty in determining orientation heterogeneity: first, the mean tilt angle measurement can be inaccurate, and second, the orientational distribution is unknown. Although this challenge has been well-known for more than a decade,[15,27,33–35] no general solutions to it have been proposed, to the best of our knowledge.

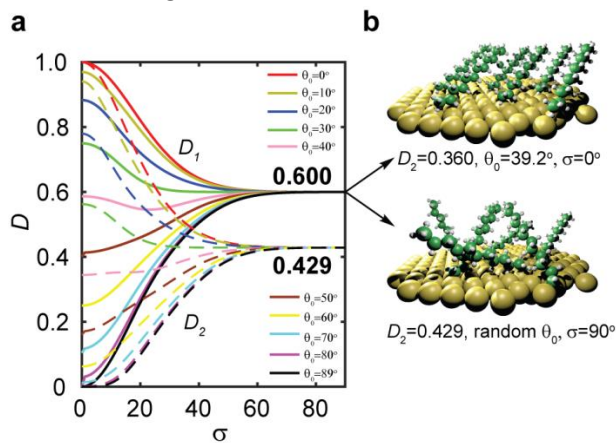

**Fig. 1** The "magic angle" challenge in determining molecular orientation and its solution by using a combination of $D_1$ and $D_2$. (a) Orientational parameters $D_1$ (solid line) and $D_2$ (dashed line) as a function of orientational distribution $\sigma$ for a series of mean tilt angles $\theta_0$. $D_1$ and $D_2$ are calculated based on a modified Gaussian function.[27] Details about this function are described in the electronic supplemental information. (b) When only $D_1$=0.600 is measured, it is unknown whether the surface has a uniform orientation distribution with $\theta_0$=39.2° or a broad orientation distribution. However, the $D_2$ is different for these two scenarios.

To determine the orientation heterogeneity, another independent quantity besides $D_1$ needs to be measured using a second spectroscopy. For instance, by combining second harmonic generation (SHG) and another linear spectroscopic method, such as angle-resolved absorbance[17], linear reflection[38,39], or polarization-resolved fluorescence detection[40–43], the orientational distribution of molecules in thin films can be determined. These methods all rely on spectroscopies that are not intrinsically surface-sensitive and, therefore, have been mainly used to study thin film samples. There have been a few attempts to determine interfacial molecular orientation using non-linear optics. Eisenthal and co-workers modified the polarization-resolved SHG by adding a circularly polarized pump pulse to create a non-equilibrium population of the surface molecules to determine the orientational distribution.[44] Also, by carefully calculating parameters related to experimental setup, Wang and co-workers showed that it is possible to extract the molecular distribution from the SHG measurement.[45]

In this work, using heterodyne two dimensional vibrational sum frequency generation (HD 2D VSFG) spectroscopy, we introduce a novel and general method to determine orientation heterogeneity and solve the "magic angle" challenge for surface science. In particular, we studied a surface catalyst system for $CO_2$ reduction as well as solar energy applications. The molecular structure and dynamics of this catalyst system have been investigated extensively with various spectroscopic methods,[46–50] but its surface molecular orientation heterogeneity has yet to be revealed, which is a critical step to better understand the surface structure-function relationship.

## Methods

### Extracting $D_1$ and $D_2$ from HD 2D VSFG

To determine surface molecular orientation heterogeneity, the key is to measure a second independent orientation parameter along with $D_1$. We used HD 2D VSFG to determine another orientation parameter $D_2$=<$\cos^5\theta$>/<$\cos\theta$>. Since $D_1$ and $D_2$ have different dependences on $\theta_0$ and $\sigma$ (Fig. 1a), a unique pair of $\theta_0$ and $\sigma$ can be determined from them.

The recently developed HD 2D VSFG spectroscopy[51–55] is the core measurement that enables surface molecular orientation heterogeneity characterization. The detailed experimental description of HD 2D VSFG spectroscopies can be found in our previous publication.[56] In brief, two vibrational coherences of a sample are created by the IR pulse train (Fig. 3a). A second coherence is subsequently upconverted nonresonantly by the narrowband 800 nm pulse. The upconverted spectra are mixed with a local oscillator for heterodyne detection, dispersed by spectrograph, and detected on a charged coupled device (CCD) camera, whose frequency axis is denoted as $\omega_3$. By scanning the time delay between the first two IR pulses ($t_1$), the upconverted spectra are modulated as a function of time, which encodes the dynamics of the first vibrational coherence. By Fourier transforming the series of spectra at different $t_1$ from time to frequency domain, the first vibrational coherence can be plotted along the $\omega_1$ axis. Overall, fourth-order vibrational susceptibilities ($\chi^{(4)}_{eff}$) are measured in HD 2D VSFG. Similar to 1D VSFG, which measures second-order susceptibilities ($\chi^{(2)}_{eff}$), these even-order nonlinear optical signals only survive in a non-centrosymmetric environment, such as interfaces. Therefore, both 1D and HD 2D VSFG spectroscopies are interface-specific vibrational spectroscopies whose signals depend on the molecular orientations.[11,16,54,57]

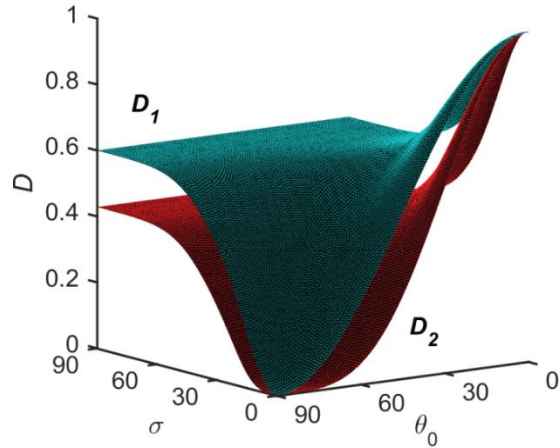

**Fig. 2** 3D surface of $D_1$ and $D_2$ as a function of mean tilt angle $\theta_0$ and orientational distribution $\sigma$. $D_1$ is plotted in blue and $D_2$ in red. The projection of these surfaces on the $D$-$\sigma$ plane is equivalent to Fig. 1a.

The key relationship that enables HD 2D VSFG spectroscopy to measure surface molecular orientation heterogeneity is that $\chi^{(2)}_{eff}$ and $\chi^{(4)}_{eff}$ can be expressed as a linear combination of $\langle\cos\theta\rangle$, $\langle\cos^3\theta\rangle$ and $\langle\cos^5\theta\rangle$ (Eq.2)[57].

$$\chi^{(2)}_{eff,i} = a_i \langle\cos\theta\rangle + b_i \langle\cos^3\theta\rangle \qquad (2)$$
$$\chi^{(4)}_{eff,i} = c_i \langle\cos\theta\rangle + d_i \langle\cos^3\theta\rangle + e_i \langle\cos^5\theta\rangle$$

These equations resulted from the Euler transformation to convert molecular frame hyperpolarizability to lab frame susceptibilities, where $a_i$, $b_i$, $c_i$, $d_i$ and $e_i$ are constants that depend on the molecular hyperpolarizabilities of the $i^{th}$ vibrational mode. The detailed derivations and expressions of $a_i$, $b_i$, $c_i$, $d_i$ and $e_i$ in terms of hyperpolarizability can be found in the electronic supplemental information. Since heterodyne 1D VSFG is simultaneously measured as heterodyne 2D VSFG is taken (see detailed data analysis in the electronic supplemental information), both $\chi^{(2)}_{eff}$ and $\chi^{(4)}_{eff}$ can be determined from the same heterodyne 2D VSFG experiments.

By taking the ratios of $\chi_{eff}$ between two vibrational modes,[58] or ratios of $\chi_{eff}$ of a single vibrational mode under different polarization combinations,[16] we obtain the key formula (Eq. 3) to extract $D_1$, $D_2$ from $\chi^{(2)}_{eff}$ and $\chi^{(4)}_{eff}$:

$$\frac{\chi^{(2)}_{eff,1}}{\chi^{(2)}_{eff,2}} = \frac{a_1 + b_1 \langle\cos^3\theta\rangle/\langle\cos\theta\rangle}{a_2 + b_2 \langle\cos^3\theta\rangle/\langle\cos\theta\rangle} = \frac{a_1 + b_1 \cdot D_1}{a_2 + b_2 \cdot D_1} \qquad (3)$$

$$\frac{\chi^{(4)}_{eff,1}}{\chi^{(4)}_{eff,2}} = \frac{c_1 + d_1 \cdot \langle\cos^3\theta\rangle/\langle\cos\theta\rangle + e_1 \cdot \langle\cos^5\theta\rangle/\langle\cos\theta\rangle}{c_2 + d_2 \cdot \langle\cos^3\theta\rangle/\langle\cos\theta\rangle + e_1 \cdot \langle\cos^5\theta\rangle/\langle\cos\theta\rangle} = \frac{c_1 + d_1 \cdot D_1 + e_1 \cdot D_2}{c_2 + d_2 \cdot D_1 + e_2 \cdot D_2}$$

$D_1$ and $D_2$ have different dependence on $\theta_0$ and $\sigma$, which is why $\theta_0$ and $\sigma$ can be solved from $D$s. The dependence of $D_1$ and $D_2$ on $\theta_0$ and $\sigma$ is summarized as 3D surface plots (Fig. 2), which we will use to demonstrate the graphic solution for searching ($\theta_0$, $\sigma$) pairs below. The $D$-$\theta_0$-$\sigma$ system can be divided into two regions: region I with $\sigma \leq 40°$ and region II with $\sigma > 40°$. In region I, because of the unique values of $D_1$ and $D_2$, the ($\theta_0$, $\sigma$) pair can be unambiguously determined. In region II, $D_1$ and $D_2$ converge to 0.600 and 0.429 asymptotically, which makes them lose the one-to-one correlation with the ($\theta_0$, $\sigma$) pair. Although the ($\theta_0$, $\sigma$) pair cannot be uniquely determined in region II, the two asymptotic number pair ($D_1$=0.600 and $D_2$=0.429) are unique signatures for broad orientational distribution. For instance, one important consequence of this asymptotic pair is solving the above-mentioned ambiguity of the "magic angle" when only $D_1$=0.600 is measured. When $D_1$ and $D_2$ are both measured, if the interfacial molecules all tilt at 39.2° with a narrow distribution, then $D_1$ should be 0.600 and $D_2$ should be 0.360 (when $\sigma$=0°, $D_2$=$D_1^2$); otherwise, with a broad angular distribution, $D_1$ and $D_2$ should be close to 0.600 and 0.429, respectively (Fig. 1b). Therefore, the orientation heterogeneity can be determined, and there is no "magic angle" ambiguity when $D_1$ and $D_2$ are measured together.

## Results

### *Determine the orientation heterogeneity of the surface catalytic monolayer*

Using Eq. 3, we can extract the orientation heterogeneity of the Re(4,4'-dicyano-2,2'-bipyridine)(CO)$_3$Cl monolayer, self-assembled on a gold slide, by measuring its HD 2D VSFG spectrum (Fig. 3c). [56] The peaks along the diagonal of the HD 2D VSFG spectrum (Fig. 3c) originate from the A'(1), A", and A'(2) modes, which reveal the orientation heterogeneity of the surface molecules. To determine $\chi^{(4)}_{eff}$, a diagonal slide of the HD 2D VSFG is taken (Fig. 3d). Heterodyne 1D VSFG spectrum is obtained directly from HD 2D VSFG raw data.

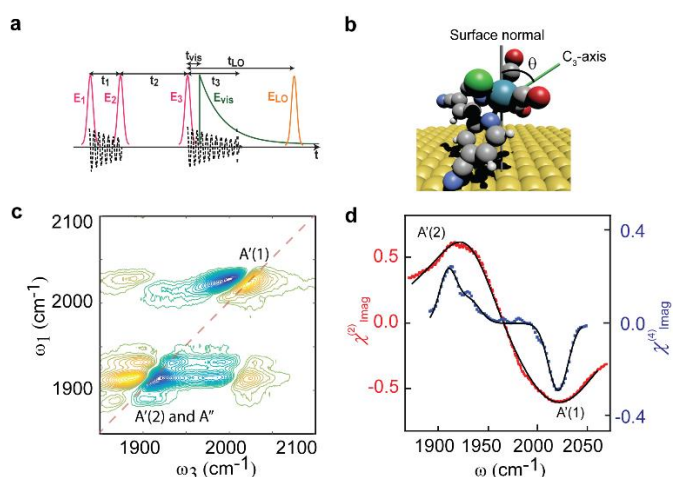

**Fig. 3** Illustration and results of heterodyne 2D VSFG spectroscopy and the surface catalysis system. (a) Heterodyne 2D VSFG pulse sequence. (b) Illustration of Re(4,4'-dicyano-2,2'-bipyridine)(CO)$_3$Cl monolayer self-assembled on a gold slide. $\theta$ is the angle between C$_3$ axis and surface normal. (c) 2D VSFG spectra of the Re complex in the carbonyl stretch region. (d) Heterodyne 1D (red dots) and diagonal cut of 2D (blue dots) VSFG spectra. The 1D spectrum has been significantly broadened by surface inhomogeneity. Solid lines represent theoretical fitting.

We determined the $\chi_{eff}$ ratios of two vibrational modes by fitting the A'(1) and A'(2) peaks in both the heterodyne 1D VSFG spectrum and the diagonal cut of the HD 2D VSFG spectrum (Fig. 3d). From this fitting, we found $\chi^{(2)}_{eff}[A'(1)]/\chi^{(2)}_{eff}[A'(2)]=-1.31\pm0.04$ and $\chi^{(4)}_{eff}[A'(1)]/\chi^{(4)}_{eff}[A'(2)]=-1.3\pm0.1$. Since all beams were held at $p$ polarization, and the Fresnel factor on gold is strong in the Z direction, we found $\chi^{(2)}_{eff} \propto \chi^{(2)}_{zzz}$ and $\chi^{(4)}_{eff} \propto \chi^{(4)}_{zzzzz}$. Using Eq. 3 and the numerical value of hyperpolarizabilities calculated by density functional theory (DFT) (B3LYP/LANL2TZ basis set, detailed methods and results in electronic supplemental information),[58] we derived the numerical relationships between $\chi^{(2)}_{zzz,1}/\chi^{(2)}_{zzz,2}$, $\chi^{(4)}_{zzzzz,1}/\chi^{(4)}_{zzzzz,2}$, and $D_1$, $D_2$. We found that $D_1=0.364\pm0.008$ and $D_2=0.14\pm0.02$.

Next, we used the measured $D_1$ and $D_2$ to extract all the qualified ($\theta_0$, $\sigma$) pairs from the $D_1$ and $D_2$ surfaces. With $D_1=0.364$ and $D_2=0.14$, we drew two planes that were parallel to the $\theta_0$-$\sigma$ plane at $D'=0.364$ and $D''=0.14$ to intersect the $D_1$ and $D_2$ surfaces (Fig. 4a and 4b), respectively. The projections of intersecting lines on the $\theta_0$-$\sigma$ plane represent the qualified ($\theta_0$, $\sigma$) pairs that have $D_1=D'$ and $D_2=D''$ (Fig. 4c), and the results agree with our previous statement that there are infinite combinations of ($\theta_0$, $\sigma$) pairs to match a single $D_1$ value. However, there is only one intersection point ($\theta_0=53°$, $\sigma=5°$) between the intersecting lines of $D_1$ and $D_2$ that represents the unique ($\theta_0$, $\sigma$) pair that satisfies both $D_1=D'$ and $D_2=D''$.

When taking into account the uncertainty of measured $D_1$ and $D_2$, the projected intersecting lines of $D_1$ and $D_2$ become stripes on the $\theta_0$-$\sigma$ plane, thus the unique intersection point turns into an intersection region, shown as the shaded region in Fig. 4d. As a result, $\theta_0$ was found to be in the region of 52~57°, while the range of $\sigma$ was different for different $\theta_0$ values. The maximum uncertainty of $\sigma$ is ±4°, when $\theta_0$ is about 53.4°. This result of $\theta_0=53°$, $\sigma=5°$ suggests that the Re-complex forms a relatively ordered layer. In this case, if assuming a narrow orientational distribution, the mean tilt angle is calculated to be 53°, same as what is determined by our new method, but the orientational distribution information is missing. Most recently, the Lian, Batista and Kubiak research groups studied the orientation of the same molecule self-assembled on a gold surface using a combination of 1D SFG and DFT simulation, and the angle between the plane of the bi-pyridine ligand and surface normal was found to be 63°.[46] However, no information about orientation heterogeneity was reported in their study.

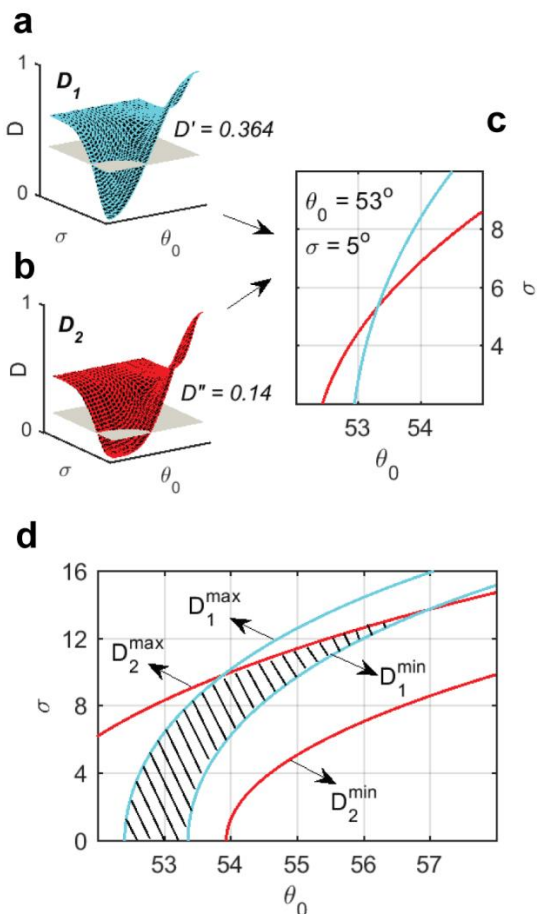

**Fig. 4** Determination of the unique ($\theta_0$, $\sigma$) pair for $D_1$=0.364 and $D_2$=0.14. Two planes that are parallel to the $\theta_0$-$\sigma$ plane are inserted at $D'$=0.364 in $D_1$ surface (a) and at $D''$=0.14 in $D_2$ surface (b). The projections of both intersecting lines on the $\theta_0$-$\sigma$ plane are plotted in (c), and the intersection point ($\theta_0$=53°, $\sigma$=5°) represents the unique ($\theta_0$, $\sigma$) pair. The region of qualified ($\theta_0$, $\sigma$) pairs for $D_1$=0.364±0.008 and $D_2$=0.14±0.02 is marked in (d).

We note that the discrepancy of the mean tilt angle determined by using our method and the traditional 1D SFG method can be large. For instance, when $D_1$=0.415 is measured and the narrow distribution is assumed, one gets 50°. However, if, with the same sample, $D_2$=0.23, then the angle would be 67° with distribution of 27°, which means that the angle determined from the narrow distribution assumption has a systematic error of 25%.

## Discussion

There are three important aspects that are influential to the orientation heterogeneity measurement. First, to properly measure surface molecular orientation heterogeneity, it is important to implement heterodyne detection, rather than homodyne detection. In heterodyne and homodyne detections, the measured 2D VSFG signals can be expressed as:[51]

$$S_{\text{heterodyne}} \approx E_{2DVSFG} * E_{LO} \propto \chi^{(4)}_{eff} \qquad (4)$$
$$S_{\text{homodyne}} \approx E_{2DVSFG} * E_{1DVSFG} \propto \chi^{(4)}_{eff} * \chi^{(2)}_{eff}$$

When heterodyne detection is used, the measured signal is proportional to $\chi^{(4)}_{eff}$, but when homodyne detection is used, the 1D VSFG signal essentially acts as a local oscillator to the heterodyne 2D VSFG signal. Therefore, the measured signal is proportional to $\chi^{(4)}_{eff} * \chi^{(2)}_{eff}$. As a result, it is difficult to disentangle these two terms and determine the $\chi^{(4)}_{eff}$ ratio from homodyne 2D VSFG.

Second, the values of $D_1$ and $D_2$ are the keys to retrieving an accurate ($\theta_0$, $\sigma$) pair, which is affected mostly by the signal-to-noise ratio (S/N) in the sum frequency generation (SFG) measurement. This becomes most apparent when $\sigma$ is relatively large, making both $D_1$ and $D_2$ converge and a small change in the $D$ values could lead to a large uncertainty. In the measurement, the S/N of HD 2D VSFG is about 20, which leads to the same order of magnitude of uncertainty in the spectra fitting mentioned above. Therefore, the experimental noise does not lead to additional uncertainty in the orientation heterogeneity measurements. We note, however, that the relation between $D_1$ and $D_2$ is restricted. For instance, a Gaussian distribution requires that $D_1^2 \leq D_2 \leq D_1$. This relation has to be satisfied if a Gaussian distribution is appropriate to describe the orientation heterogeneity of the interfacial molecules; otherwise, other models have to be proposed.

Third, another source of uncertainty comes from the value of hyperpolarizability. Here, we used high level DFT calculations to determine the hyperpolarizability, which is a common approach used in Raman and 1D SFG spectroscopic studies.[59,60] To evaluate the dependence of the ($\theta_0$, $\sigma$) pair on the basis set and the resulting hyperpolarizability, we tested basis sets at different levels in the DFT calculation. We found the calculated hyperpolarizability converges to a level where the variation of hyperpolarizability resulting from different basis sets does not cause significant changes to the resulted value of the ($\theta_0$, $\sigma$) pair (see electronic supplemental information). Another alternative for determining hyperpolarizability is to derive it from the experimental Raman depolarization ratio.[16] However, this method is limited to symmetric vibrational modes with perfect $C_{3v}$ or $C_\infty$ symmetry. Since the vibrational modes of Re-complex studied here do not have a perfect $C_{3v}$ symmetry, the Raman depolarization approach cannot be directly applied for our study. Nevertheless, a comparison between the experimental measured and the DFT calculated depolarization ratios could indicate the accuracy of the hyperpolarizability calculation. In our work, the results agree with each other reasonably well (see electronic supplemental information for more details). Further investigations of determining hyperpolarizabilities of complex molecules that are lack of rigorous symmetries is important, in order to accurately determine molecular heterogeneity, but it is out of the scope of this initial report.

## Conclusions

The method in this work can be applied broadly to determine the orientation heterogeneity of interfacial molecules in systems consisting of water, solid state materials and biological molecules, many of which have been investigated by HD 2D VSFG spectroscopy.[52–55,61–65] Therefore, the orientation analysis presented here can be directly applied to these spectral regimes. In addition, since only the diagonal cuts of 2D VSFG are analyzed to obtain the orientation heterogeneity, in principle, the same method can be applied to IR pump-SFG probe experiments[50,66] when there is no coupling between vibrational modes. This extension enables the orientation heterogeneity measurement by a more established technique than HD 2D SFG.

In summary, we demonstrate that the orientation heterogeneity of molecules at interfaces can be determined using HD 2D VSFG. In particular, we studied the monolayer of Re(4,4'-dicyano-2,2'-bipyridine)(CO)$_3$Cl on a gold electrode and found that it forms a fairly ordered monolayer. This new advancement solves the long-standing "magic angle" challenge seen in 1D VSFG spectroscopy. With the growing popularity of HD 2D VSFG spectroscopy in the surface science community,[51–57,62–65,67–69] this new method will contribute significantly in determining the molecular conformations of interfaces in solid state materials, water, biological membranes and many other important interfaces.

## Acknowledgements

We thank M. Stone and Prof. M. Tauber for their help in measuring Raman depolarization ratio of the Re-complex and inspiring discussion on the related topics. We acknowledge M. L. Clark and Prof. C. P. Kubiak for providing the surface catalyst sample. We thank M. Riera, P. Bajaj and Prof. F. Paesani for helping us set up the DFT calculation. This project is supported by The Defense Advanced Research Projects Agency (government grant number D15AP000107).

## References

1   K. Brandt, M. E. Chiu, D. J. Watson, M. S. Tikhov and R. M. Lambert, *J. Am. Chem. Soc.*, 2009, **131**, 17286–17290.
2   D. G. G. McMillan, S. J. Marritt, G. L. Kemp, P. Gordon-Brown, J. N. Butt and L. J. C. Jeuken, *Electrochim. Acta*, 2013, **110**,

79–85.
3   R. Barbey, L. Lavanant, D. Paripovic, N. Schuwer, C. Sugnaux, S. Tugulu and H.-A. Klok, *Chem. Rev.*, 2009, **109**, 5437–5527.
4   A. E. Jailaubekov, A. P. Willard, J. R. Tritsch, W.-L. Chan, N. Sai, R. Gearba, L. G. Kaake, K. J. Williams, K. Leung, P. J. Rossky and X.-Y. Zhu, *Nat. Mater.*, 2013, **12**, 66–73.
5   G. Wu and K. Y. C. Lee, *Langmuir*, 2009, **25**, 2133–2139.
6   K. L. H. Lam, Y. Ishitsuka, Y. Cheng, K. Chien, A. J. Waring, R. I. Lehrer and K. Y. C. Lee, *J. Phys. Chem. B*, 2006, **110**, 21282–21286.
7   Y. R. Shen, *Nature*, 1989, **337**, 519–525.
8   R. Superfine, J. Y. Huang and Y. R. Shen, *Phys. Rev. Lett.*, 1991, **66**, 1066–1069.
9   Q. Du, R. Superfine, E. Freysz and Y. R. Shen, *Phys. Rev. Lett.*, 1993, **70**, 2313–2316.
10  K. B. Eisenthal, *Chem. Rev.*, 1996, **96**, 1343–1360.
11  H.-F. Wang, L. Velarde, W. Gan and L. Fu, *Annu. Rev. Phys. Chem.*, 2015, **66**, 189–216.
12  F. M. Geiger, *Annu. Rev. Phys. Chem.*, 2009, **60**, 61–83.
13  E. C. Y. Yan, Z. Wang and L. Fu, *J. Phys. Chem. B*, 2015, **119**, 2769–2785.
14  Z. Li, C. N. Weeraman and J. M. Gibbs-Davis, *ChemPhysChem*, 2014, **15**, 2247–2251.
15  S. Ye, K. T. Nguyen, A. P. Boughton, C. M. Mello and Z. Chen, *Langmuir*, 2010, **26**, 6471–6477.
16  H.-F. Wang, W. Gan, R. Lu, Y. Rao and B.-H. Wu, *Int. Rev. Phys. Chem.*, 2005, **24**, 191–256.
17  G. Simpson, S. Westerbuhr and K. Rowlen, *Anal. Chem.*, 2000, **72**, 887–898.
18  R. R. Naujok, D. A. Higgins, D. G. Hanken and M. Robert, *J. Chem. Soc. Faraday Trans.*, 1995, **91**, 1411–1420.
19  S. Roy, P. a Covert, W. R. FitzGerald and D. K. Hore, *Chem. Rev.*, 2014, **114**, 8388–8415.
20  X. Chen, M. L. Clarke, J. Wang and Z. Chen, *Int. J. Mod. Phys. B*, 2005, **19**, 691–713.
21  A. G. F. De Beer and S. Roke, *J. Chem. Phys.*, 2010, **132**, 1–7.
22  Z. Li, C. N. Weeraman, M. S. Azam, E. Osman and J. M. Gibbs-Davis, *Phys. Chem. Chem. Phys.*, 2015, **17**, 12452–12457.
23  A. M. Buchbinder, E. Weitz and F. M. Geiger, *J. Am. Chem. Soc.*, 2010, **132**, 14661–14668.
24  T. Weidner, J. S. Apte, L. J. Gamble and D. G. Castner, *Langmuir*, 2010, **26**, 3433–3440.
25  L. Fu, J. Liu and E. C. Y. Yan, *J. Am. Chem. Soc.*, 2011, **133**, 8094–8097.
26  X. Zhuang, P. Miranda, D. Kim and Y. Shen, *Phys. Rev. B*, 1999, **59**, 12632–12640.
27  G. J. Simpson and K. L. Rowlen, *J. Am. Chem. Soc.*, 1999, **121**, 2635–2636.
28  M. L. Clarke, C. Chen, J. Wang and Z. Chen, *Langmuir*, 2006, **22**, 8800–8806.
29  J. Wang, Z. Paszti, M. a. Even and Z. Chen, *J. Am. Chem. Soc.*, 2002, **124**, 7016–7023.
30  Gary P. Harp, H. Rangwalla, M. S. Yeganeh and A. Dhinojwala, *J. Am. Chem. Soc.*, 2003, **125**, 11283–11290.
31  I. Rocha-Mendoza, D. R. Yankelevich, M. Wang, K. M. Reiser, C. W. Frank and A. Knoesen, *Biophys. J.*, 2007, **93**, 4433–4444.
32  K. T. Nguyen, S. V Le Clair, S. Ye and Z. Chen, *J. Phys. Chem. B*, 2009, **113**, 12169–12180.
33  H. Zhu and A. Dhinojwala, *Langmuir*, 2015, **31**, 6306–6313.
34  X. Chen, B. Minofar, P. Jungwirth and H. C. Allen, *J. Phys. Chem. B*, 2010, **114**, 15546–15553.
35  Z. Li, C. N. Weeraman and J. M. Gibbs-Davis, *J. Phys. Chem. C*, 2014, **118**, 28662–28670.
36  S. Watcharinyanon, C. Puglia, E. Göthelid, J. E. Bäckvall, E. Moons and L. S. O. Johansson, *Surf. Sci.*, 2009, **603**, 1026–1033.
37  M. Friedrich, G. Gavrila, C. Himcinschi, T. U. Kampen, a Y. Kobitski, H. Méndez, G. Salvan, I. Cerrilló, J. Méndez, N. Nicoara, a M. Baró and D. R. T. Zahn, *J. Phys. Condens. Matter*, 2003, **15**, S2699–S2718.
38  S. Yamaguchi, H. Hosoi, M. Yamashita, P. Sen and T. Tahara, *J. Phys. Chem. Lett.*, 2010, **1**, 2662–2665.
39  A. Kundu, H. Watanabe, S. Yamaguchi and T. Tahara, *J. Phys. Chem. C*, 2013, **117**, 8887–8891.
40  M. A. Bos and J. M. Kleijn, *Biophys. J.*, 1995, **68**, 2566–72.
41  P. L. Edmiston, J. E. Lee, S.-S. Cheng and S. S. Saavedra, *J. Am. Chem. Soc.*, 1997, **119**, 560–570.
42  A. Tronin, J. Strzalka, X. Chen, P. L. Dutton and J. K. Blasie, *Langmuir*, 2000, **16**, 9878–9886.
43  A. F. Runge, S. S. Saavedra and S. B. Mendes, *J. Phys. Chem. B*, 2006, **110**, 6721–6731.
44  Y. Rao, S.-Y. Hong, N. J. Turro and K. B. Eisenthal, *J. Phys. Chem. C*, 2011, **115**, 11678–11683.
45  Y. Rao, Y. Tao and H. Wang, *J. Chem. Phys.*, 2003, **119**, 5226–5236.
46  M. L. Clark, B. Rudshteyn, A. Ge, S. A. Chabolla, C. W. Machan, B. T. Psciuk, J. Song, G. Canzi, T. Lian, V. S. Batista and C. P. Kubiak, *J. Phys. Chem. C*, 2016, **120**, 1657–1665.
47  D. E. Rosenfeld, Z. Gengeliczki, B. J. Smith, T. D. P. Stack and M. D. Fayer, *Science (80-. ).*, 2011, **334**, 634–639.
48  C. Yan, R. Yuan, W. C. Pfalzgraff, J. Nishida, L. Wang, T. E. Markland and M. D. Fayer, *PNAS*, 2016, **113**, 4929–4934.
49  W. Xiong, J. E. Laaser, P. Paoprasert, R. A. Franking, R. J. Hamers, P. Gopalan and M. T. Zanni, *J. Am. Chem. Soc.*, 2009, **131**, 18040–18041.

50  C. L. Anfuso, A. M. Ricks, W. Rodríguez-Cordoba and T. Lian, *J. Phys. Chem. C*, 2012, **116**, 26377–26384.
51  W. Xiong, J. E. Laaser, R. D. Mehlenbacher and M. T. Zanni, *Proc. Natl. Acad. Sci.*, 2011, **108**, 20902–20907.
52  K. Inoue, S. Nihonyanagi, P. C. Singh, S. Yamaguchi and T. Tahara, *J. Chem. Phys.*, 2015, **142**, 212431.
53  P. C. Singh, S. Nihonyanagi, S. Yamaguchi and T. Tahara, *J. Chem. Phys.*, 2012, **137**, 094706.
54  J. E. Laaser, D. R. Skoff, J. J. Ho, Y. Joo, A. L. Serrano, J. D. Steinkruger, P. Gopalan, S. H. Gellman and M. T. Zanni, *J. Am. Chem. Soc.*, 2014, **136**, 956–962.
55  Z. Zhang, L. Piatkowski, H. J. Bakker and M. Bonn, *Nat. Chem.*, 2011, **3**, 888–893.
56  J. Wang, M. L. Clark, Y. Li, C. L. Kaslan, C. P. Kubiak and W. Xiong, *J. Phys. Chem. Lett.*, 2015, **6**, 4204–4209.
57  J. E. Laaser and M. T. Zanni, *J. Phys. Chem. A*, 2013, **117**, 5875–5890.
58  H. Wu, W.-K. Zhang, W. Gan, Z.-F. Cui and H.-F. Wang, *J. Chem. Phys.*, 2006, **125**, 133203.
59  H. M. Chase, B. T. Psciuk, B. L. Strick, R. J. Thomson, V. S. Batista and F. M. Geiger, *J. Phys. Chem. A*, 2015, **119**, 3407–3414.
60  L. Fu, D. Xiao, Z. Wang, V. S. Batista and E. C. Y. Yan, *J. Am. Chem. Soc.*, 2013, **135**, 3592–3598.
61  J. Bredenbeck, A. Ghosh, M. Smits and M. Bonn, *J. Am. Chem. Soc.*, 2008, **130**, 2152–2153.
62  J. Bredenbeck, A. Ghosh, H.-K. Nienhuys and M. Bonn, *Acc. Chem. Res.*, 2009, **42**, 1332–1342.
63  P. C. Singh, S. Nihonyanagi, S. Yamaguchi and T. Tahara, *J. Chem. Phys.*, 2013, **139**, 161101.
64  Z. Zhang, L. Piatkowski, H. J. Bakker and M. Bonn, *J. Chem. Phys.*, 2011, **135**, 021101.
65  C. S. Hsieh, M. Okuno, J. Hunger, E. H. G. Backus, Y. Nagata and M. Bonn, *Angew. Chemie - Int. Ed.*, 2014, **53**, 8146–8149.
66  A. Eftekhari-Bafrooei and E. Borguet, *J. Am. Chem. Soc.*, 2009, **131**, 12034–12035.
67  A. Ghosh, J.-J. Ho, A. L. Serrano, D. R. Skoff, T. Zhang and M. T. Zanni, *Faraday Discuss.*, 2015, **177**, 493–505.
68  J.-J. Ho, D. R. Skoff, A. Ghosh and M. T. Zanni, *J. Phys. Chem. B*, 2015, **119**, 10586–10596.
69  Y. Li, J. Wang, M. L. Clark, C. P. Kubiaka and W. Xiong, *Chem. Phys. Lett.*, 2016, **650**, 1–6.